\documentclass[prl
,twocolumn
,showpacs
,amsmath
,aps]{revtex4}
\usepackage{natbib}
\usepackage{graphicx}
\begin{document}
\title{Transition from antibunching to bunching in cavity QED}
\author{Markus Hennrich}
\author{Axel Kuhn}
\author{Gerhard Rempe}
\affiliation{Max-Planck-Institut f\"{u}r Quantenoptik, Hans-Kopfermann-Str.\,1, D-85748 Garching, Germany}
\date{\today}
\begin{abstract}
The photon statistics of the light emitted from an atomic ensemble into a single field mode of an optical cavity is investigated as a function of the number of  atoms. The light is produced in a Raman transition driven by a pump laser and the cavity vacuum [M.~Hennrich et al., Phys.\,Rev.\,Lett.\,{\bf 85}, 4672 (2000)], and a recycling laser is employed to repeat this process continuously. 
 For weak driving, a smooth transition from antibunching to bunching is found for about one intra-cavity atom. Remarkably, the bunching peak develops within the antibunching dip. For saturated driving and a growing number of atoms, the bunching amplitude decreases and the bunching duration increases, indicating the onset of Raman lasing. 
\end{abstract}


\maketitle

The photon statistics of light reveals whether it originates from a classical thermal or coherent source, or from a quantum source like a single atom. These sources can be distinguished by their intensity correlation function, $g^{(2)}(\tau)$ \cite{Scully97}. Classical light fulfils the Schwarz inequality, $g^{(2)}(0) \ge g^{(2)}(\tau)$, whereas light that violates this inequality must be described by the laws of quantum physics. First experiments demonstrating antibunching, $g^{(2)}(0) < g^{(2)}(\tau)$,  were performed with a weak beam of atoms \cite{Kimble77}.  Limitations imposed by the number fluctuations of the atomic beam \cite{Carmichael78} were later eliminated by using a single ion \cite{Diedrich87}, atom \cite{Gomer98,McKeever03:2}, molecule \cite{Martini96}, quantum dot \cite{Michler00},  or  color centre \cite{Kurtsiefer00,Beveratos01}. The deterministic control of the nonclassical light radiated by a single emitter is now a very active field of research \cite{Kim99,Lounis00,Michler00:2,Santori02,Kuhn02,Beveratos02:2,McKeever04}, with many interesting applications, e.g., in quantum information processing.

Classical bunching, $g^{(2)}(0) > g^{(2)}(\tau)$,  has been observed in the fluorescence of a large number of independently radiating atoms as early as 1956 \cite{Hanbury56}, and has regained new interest in the context of cold-atom physics  \cite{Jurczak96,Bali96}. A smooth transition between antibunching and bunching is expected if the number of contributing atoms gradually increases. Surprisingly, such a transition has not been observed so far. The reason for this is that,  to detect antibunching, a good photon collection efficiency, and thus a large solid angle, is essential to obtain a sufficiently large signal. In contrast, to observe bunching, spatial coherence of the detected light is required. For a distributed ensemble of atoms, this can only be realized by monitoring a single (diffraction-limited) light mode covering a small solid angle. Obviously, these two requirements contradict each other \cite{Narducci96}, making the experiment difficult in the interesting regime of just a few radiating particles.

In the work presented here, all emitters are coupled to a single mode in a high-finesse optical cavity. Only the light in this mode is investigated, so that spatial coherence is granted. At the same time, the enhanced spontaneous emission into the cavity mode gives a good photon collection efficiency. Moreover, the experiment is performed in a regime where an emitted photon leaves the cavity before affecting other atoms. This results in a small interaction between different atoms and therefore small collective effects that otherwise would lead to a novel photon statistics \cite{Rempe91,Mielke98,Smith02}. Therefore all requirements to observe the transition between antibunching and bunching with one-and-the-same experimental setup are fulfilled. In fact, we find that with increasing number of atoms, a strong bunching peak (demonstrating the wave character of the light) develops inside the antibunching minimum at $\tau=0$ (characterising the particle nature of the light).

\begin{figure}
  \begin{center}
    \includegraphics[width=8.5cm]{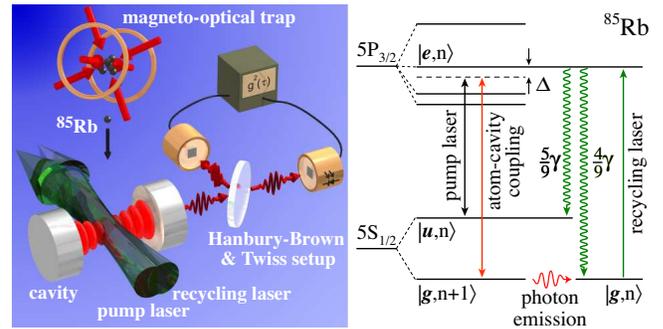}\\
  \end{center}
  \caption{Scheme of the experiment. \textbf{left:} Setup: Atoms are released from a magneto-optical trap and fall through a cavity $20\,$cm below with a
velocity of $2\,$m/s. Each atom interacts with the TEM$_{00}$ mode of the cavity for about $20\,\mu$s and is at the same time exposed to pump and recycling laser beams. The light emitted from the cavity is registered by a pair of photo diodes.
\textbf{right:} Relevant levels and transitions in $^{85}$Rb. The atomic states labelled $|u\rangle$, $|e\rangle $ and $|g\rangle $ are involved in the Raman process, and the states $|n\rangle $ and $|n+1\rangle $ denote the photon number in the cavity.}
  \label{fig:setup}
\end{figure}

Figure \ref{fig:setup} illustrates the setup. A cloud of $^{85}$Rb atoms
%, prepared in state $|u\rangle \equiv |5S_{1/2}(F=3)\rangle$, 
is released from a magneto-optical trap (MOT) and falls through a 1\,mm long optical cavity of finesse $F=60000$. The density of the cloud, and therefore the average number of atoms, $\bar{N}$, that simultaneously interact with the TEM$_{00}$ mode of the cavity, is freely adjustable by the loading time of the trap between $\bar{N}=0$ and $\bar{N}\approx 140$. While the atoms fall through the cavity, they are exposed to two laser beams. The pump laser continuously drives the transition between state $|u\rangle \equiv |5S_{1/2}(F=3)\rangle$ and the excited state $|e\rangle\equiv |5P_{3/2}(F=3)\rangle$ with Rabi frequency $\Omega_P$, while the cavity couples $|e\rangle$ to the other hyperfine ground state, $|g\rangle \equiv |5S_{1/2}(F=2)\rangle$. Both fields are detuned by an amount $\Delta$ from the respective atomic transition so that they resonantly drive a Raman transition between $|u\rangle$ and $|g\rangle$ which also changes the photon number by one. At the same time,  a recycling laser of Rabi frequency $\Omega_R$ resonantly drives the transition from $|g\rangle$ to $|e\rangle$, from where the atoms decay back to state $|u\rangle$.  This closes the excitation loop and enables each atom to emit more than one photon on its way through the cavity. Due to the continuous driving,  the Raman transitions are stochastic in contrast to the STIRAP process reported in \cite{Hennrich00,Kuhn02,Kuhn02:2,Hennrich03,Kuhn03}. The dynamics of the system is determined by $(g_\text{max}, \kappa, \gamma, \Omega_P, \Omega_R, \Delta) =  2\pi \times (2.5, 1.25, 3.0, 7.6, 3.3, -20)\,$MHz, where $g_\text{max}$ is the cavity-induced coupling between states $|e,0\rangle$ and $|g,1\rangle$ for an atom optimally coupled to the cavity, and $\kappa$ and $\gamma$ are the field and polarization decay rates of the cavity and the atom, respectively. The maximum recycling rate is achieved when the transition between $|g\rangle$ and $|e\rangle$ is strongly saturated. In this case, both levels are equally populated leading to a recycling rate of $R_\text{max}=\frac{5}{9} \gamma  = 2\pi \times 1.7\,$MHz, where $\frac{5}{9}$ is the average branching ratio for a decay from $|e\rangle$ to $|u\rangle$. Therefore the recycling is always slower than the decay of the cavity excitation, $2\kappa$. For the above value of $\Omega_{R}$, the  recycling rate is even smaller, so that  non-classical antibunching can be observed. The maximum effective Rabi frequency of the Raman process, $\Omega_\text{eff}=g_\text{max} \Omega_P /\Delta=2\pi \times 0.95\,$MHz, is also smaller than the cavity decay rate. Therefore the system is overdamped and shows no Rabi oscillations, i.e. both the reabsorption of emitted photons and the cavity-mediated interaction between different atoms are negligible. The cavity decay is mainly caused by the 100 ppm  transmittance of one of the mirrors. Photons leave the cavity through this output coupler with a probability of 90\%. They are detected by two avalanche photo diodes with 50\% quantum efficiency that are placed at the output ports of a beam splitter. They form a Hanbury-Brown and Twiss setup to measure the $g^{(2)}(\tau)$ intensity correlation function of the emitted light.

\begin{figure}
  \begin{center}
    \includegraphics[width=8.5cm]{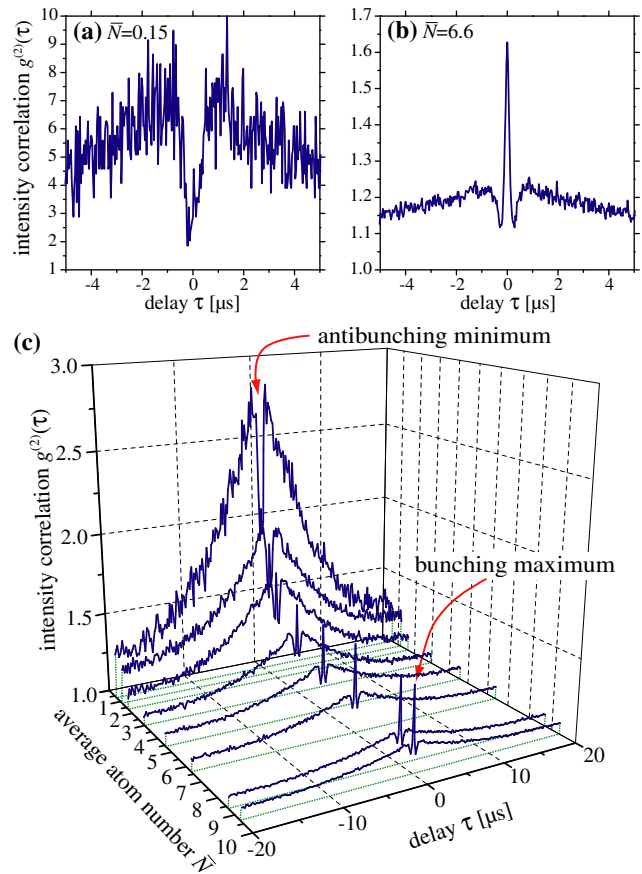}\\
  \end{center}
  \caption{Intensity correlation, $g^{(2)}(\tau)$, as a function of the detection time  delay, $\tau$, for different values of the atom flux and weak driving. For an average atom number around  one, a transition from antibunching ($g^{(2)}(0)<g^{(2)}(\tau)$) to bunching  ($g^{(2)}(0)>g^{(2)}(\tau)$) is observed. To adjust the atom flux, we load the trap  between $20\,$ms and $2.5\,$s. For each of the  experimental traces, we load and release atoms from the MOT $500$ times and register  photons during $\delta t=8\,$ms while the atom cloud traverses the cavity.  The intensity correlation, $g^{(2)}(\tau)$, is calculated from the recorded photodetection times.  }
  \label{fig:graf3d}
\end{figure}

Fig.~\ref{fig:graf3d} shows $g^{(2)}(\tau)$ for different settings of the atom flux. For an average atom number below one, $\bar{N}=0.15$, it shows a non-classical behavior with strong antibunching, i.e.~$g^{(2)}(0) < g^{(2)}(\tau)$ in Fig.~\ref{fig:graf3d} (a). Note that sub-Poissonian light with $g^{(2)}(0)<1$ is not observed because the Poissonian atom statistics is mapped to the photon statistics. When  the atom flux is increased to $\bar{N}>1$,  a transition to bunching, i.e.~$g^{(2)}(0)>g^{(2)}(\tau)$, is observed, see Fig.~\ref{fig:graf3d}(b) and (c). 

This transition from non-classical light for $\bar{N}<1$ to classical light for $\bar{N}>1$ is well explained by a model that describes an ensemble of independent emitters where the common electric field of all atoms is the sum of the individual fields \cite{Carmichael78}. Neglecting correlations between different atoms and detector noise (which is small compared to the mean count rate), and provided the atom distribution is Poissonian with an average atom number $\bar{N}$ \footnote{The  MOT-loading time leading to $\bar{N}=1$ is determined using  $g^{(2)}(1\,\mu$s$,\bar{N}= 1) \approx 2$ under the assumptions that $g_A^{(2)}(\tau\ge 1\,\mu$s$)=1$ and $g_{A}^{(1)}(\tau\ge 1\,\mu$s$)\approx 0$, with $f(1\,\mu$s$) \approx 1$. Relative to this calibration, the average atom number for different atom flux is determined by  laser-induced fluorescence.}, this leads to
\begin{equation}\label{eq:ICorrIA}
g^{(2)}(\tau) =
1+|f(\tau) g_A^{(1)}(\tau)|^2+f(\tau)\frac{g_A^{(2)}(\tau)}{\bar{N}}
\end{equation}
which consists of three different components:
\textbf{(1)} 
The constant term 1 stems from photons that are independently emitted by different atoms, i.e. it reflects the atom statistics which is directly mapped to the light.
\textbf{(2)}
The bunching term, $f(\tau)^2|g_A^{(1)}(\tau)|^2$, with $g_{A}^{(1)}(\tau)$ the auto-correlation function of the electric field emitted by one atom and $f(\tau)$ given below \cite{Kimble78}, results from the beating of the light emitted by different atoms. Constructive or destructive interference leads to a fluctuating intensity \cite{Scully97}. If a photon is detected, constructive interference is likely and the probability for a second photodetection is increased. The opposite holds true for destructive interference. This effect demonstrates the wave character of the light. The interference and the correlated behavior vanish if the two photodetections are separated by more than the coherence time. Therefore the bunching contribution decreases with the square of $g_A^{(1)}(\tau)$, whose $\frac{1}{\mathrm{e}}$-decay defines the coherence time, $\tau_c$. Note that this contribution does not depend on the number of atoms and therefore persists for very high atom flux.
\textbf{(3)}
The antibunching term, $f(\tau){g_A^{(2)}(\tau)}/{\bar{N}}$, with $g_{A}^{(2)}(\tau)$ the single-atom intensity correlation function, is attributed to the photons emitted from an individual atom. After a photon emission, the atom must be recycled to state $|u\rangle$ before it can emit another photon. The time-lag between these photons leads to the antibunching minimum at $\tau=0$. This demonstrates the particle nature of light. Due to the statistical nature of the recycling, photons are uncorrelated for $|\tau|\rightarrow\infty$. Therefore the intensity correlation function of an individual atom, $g_A^{(2)}(\tau)$, reaches 1 for $|\tau|\rightarrow\infty$. However, only photons emitted from one-and-the-same atom during its limited  interaction time with the cavity contribute to $g_A^{(2)}(\tau)$. The empirically found envelope function, $f(\tau)=\exp(-(|\tau|/\tau_i)^{1.3})$, with $\tau_{i}=7.1\,\mu$s characterizing the interaction time, corrects for this effect. Note that the antibunching term scales with the inverse average atom number, $1/\bar{N}$, and therefore vanishes for large $\bar{N}$.

\begin{figure}
  \begin{center}
    \includegraphics[width=8cm]{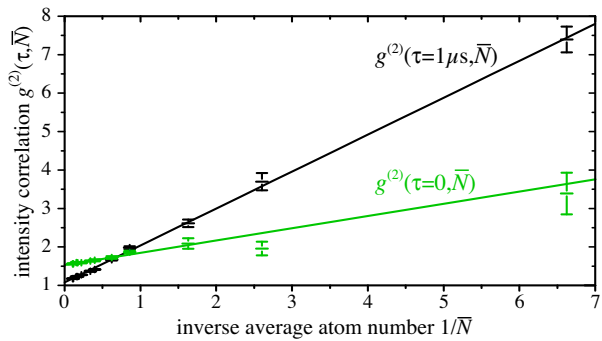}\\
  \end{center}
  \caption{Intensity correlation, $g^{(2)}(\tau,\bar{N})$, at $\tau=0$ and  $\tau=1\,\mu$s as a function of the inverse average atom number, $1/\bar{N}$. }
  \label{fig:g2scaling}
\end{figure}

The three contributions explain the observed transition from antibunching to bunching with increasing atom number. The characteristics are directly visible in Fig.~\ref{fig:graf3d}(c): the antibunching contribution for $\bar{N}<1$ vanishes with increasing atom number while the bunching contribution does not change.

Now, we present a detailed comparison of this model with the experimental results by fitting an equation of the type $1+A+B/\bar{N}$ to the experimental data. Hyperbolic scaling of the antibunching contribution is expected as a function of $\bar{N}$, while the other contributions should not be sensitive to the atom number. Fig.~\ref{fig:g2scaling} shows that this behavior is indeed found in the experiment. Two traces of $g^{(2)}(\tau, \bar{N})$ are shown as a function of $1/\bar{N}$ for $\tau=0$ and $\tau=1\,\mu$s, i.e.~along the maxima of the bunching term and the antibunching contribution, respectively. Obviously,  both traces scale like $1/\bar{N}$. The slope is proportional to the amplitude of the antibunching component whereas the  value for $1/\bar{N}\rightarrow 0$ reflects the constant offset due to the other two contributions, (1) and (2) from above. As expected, the trace along $\tau=1\,\mu$s has a large slope, i.e. a significant ${1}/{\bar{N}}$ contribution from the antibunching term, and reaches a value very close to 1 for $1/\bar{N}\rightarrow 0$, since the bunching contribution is negligible for large $\tau$. In contrast, the trace along $\tau=0$ has a much smaller slope. However, the slope  does not vanish as one would expect for usual resonance fluorescence of a single atom with $g_{A}^{(2)}(0)=0$ \cite{Carmichael76}. Instead, we obtain $g_A^{(2)}(0)=0.32 \pm 0.01$. This reflects the small but non-negligible probability that a single atom emits a second photon before the first photon has left the cavity. Moreover, in the limit $1/\bar{N}\rightarrow 0$,  the value measured in our experiment is $g^{(2)}(0)=1.53 \pm 0.01$. This deviates slightly from the expectations for independently emitting atoms, which should give rise to $g^{(2)}(0)=f(0)^2|g_A^{(1)}(0)|^2+1=2$ since $g_A^{(1)}(0)\equiv 1$ and $f(0)=1$. We attribute this deviation from the value 2 to cavity-mediated atom-atom interactions, as discussed now.

\begin{figure}
  \begin{center}
    \includegraphics[width=8.5cm]{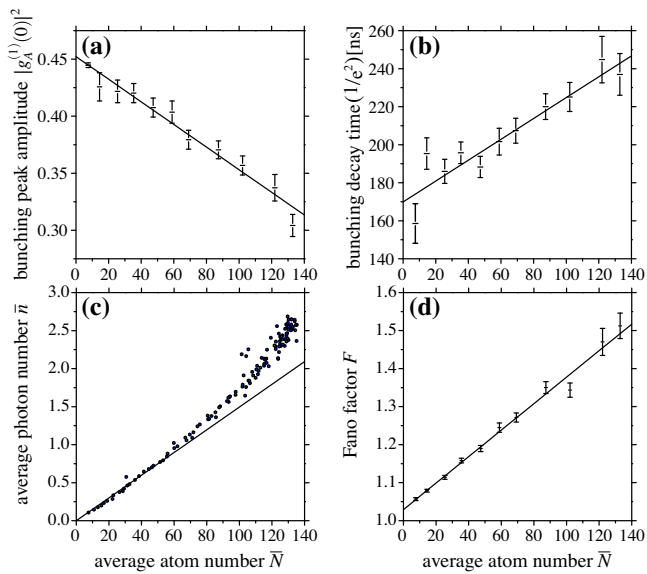}\\
  \end{center}
  \caption{Collective effects for saturated driving.
  \textbf{(a)} Amplitude of the bunching peak, $|g_{A}^{(1)}(0)|^2$, as a function of the atom number.
  \textbf{(b)} Decay time of the bunching contribution indicating the coherence time $\tau_c$.  It increases with $\bar{N}$, like for a laser at threshold.
  \textbf{(c)} Average photon number in the cavity.  The nonlinear behavior indicates light amplification by stimulated emission.
  \textbf{(d)} The Fano factor, $F$, increases with $\bar{N}$ but does not reach a maximum.}
  \label{fig:collective}
\end{figure}

From the finite duration of the bunching contribution, (2), we calculate a coherence time of the emitted light of $\tau_c=148\,$ns for $\bar{N}\le 10$. This is close to  the decay time of the cavity field, $\kappa^{-1} = 127\,$ns,  and therefore indicates that the assumption of having independent emitters is justified for weak driving.
%
%\begin{comment}
%In fact a different definition of the coherence time \cite[chap.4.3.3]{Mandel95} swaps the proportions, since the bunching maximum does not decay exponentially, as would be expected for a Lorentzian spectrum, but has a gaussian form. This form hints to an inhomogeneous line broadening of the emitted photons, e.g. due to different frequency shifts of the hyperfine sub-levels caused by the magnetic guidance field.
%\end{comment}
%
In fact, Fig.~\ref{fig:collective} shows that collective effects become only apparent if  the average atom number is increased by one order of magnitude and the Rabi frequencies are raised to $(\Omega_P, \Omega_R) = 2\pi \times  (20, 12)\,$MHz. In this case, the amplitude of the bunching term decreases as a function of $\bar{N}$ from $0.45$ (at $\bar{N}=7.5$) to $0.30$ (at $\bar{N}=130$), while its width increases from $160\,$ns to $240\,$ns. This indicates that the coherence length of the light increases with the number of atoms, as is expected for a laser at threshold. However, a kink in the average photon number, that would signal the threshold of a conventional laser, is not observed here (Fig.~\ref{fig:collective}~(c)). Only a moderate amplification is visible. For a small atom number, the average number of photons per atom circulating in the cavity is $\bar{n}/\bar{N}=0.015$, whereas for $\bar{N}=130$, a value of $\bar{n}/\bar{N}=0.02$ is reached, i.e.~the light circulating in the cavity is amplified. However, the Fano factor of the emitted light, $F\equiv {\Delta n^2} / {\bar{n}} = \bar{n} (g^{(2)}(0)-1) + 1$, increases linearly in the entire range (Fig.~\ref{fig:collective}~(d)). No maximum is observed that would indicate a laser threshold. This is consistent with a numerical simulation performed for our experimental parameters. To achieve laser operation with a few or possibly only one atom, a recycling rate significantly larger than the cavity decay rate is mandatory, so that the photons remain long enough in the cavity to stimulate  further emissions. A first step into this direction with a large atom-cavity coupling has recently been made by McKeever et al. with a single trapped atom \cite{McKeever03:2}.  In this work, the observed antibunching was attributed to  thresholdless lasing.

In conclusion, we have observed the transition from antibunching to bunching in the light emitted from a high-finesse cavity as a function of the average number of atoms coupled to the cavity. The cavity decay determines the fastest time scale, so that the atoms can be considered as independent emitters. Only for average atom numbers above $\bar{N} \approx 60$, cavity-mediated atom-atom interactions lead to collective effects, causing an amplification of the light circulating in the cavity and therefore an increase of the coherence time. By sufficiently increasing the cavity finesse, it should be possible to realize a single-atom laser producing a coherent light field with  time-independent intensity correlation function.

\begin{acknowledgements}
We thank B.\,W.\,Shore for stimulating discussions. This work was supported by the focused research program `Quantum Information Processing' and the SFB 631 of the Deutsche Forschungsgemeinschaft, and by the European Union through the IST (QGATES) and IHP (CONQUEST) programs.
\end{acknowledgements}


\begin{thebibliography}{29}
\expandafter\ifx\csname natexlab\endcsname\relax\def\natexlab#1{#1}\fi
\expandafter\ifx\csname bibnamefont\endcsname\relax
  \def\bibnamefont#1{#1}\fi
\expandafter\ifx\csname bibfnamefont\endcsname\relax
  \def\bibfnamefont#1{#1}\fi
\expandafter\ifx\csname citenamefont\endcsname\relax
  \def\citenamefont#1{#1}\fi
\expandafter\ifx\csname url\endcsname\relax
  \def\url#1{\texttt{#1}}\fi
\expandafter\ifx\csname urlprefix\endcsname\relax\def\urlprefix{URL }\fi
\providecommand{\bibinfo}[2]{#2}
\providecommand{\eprint}[2][]{\url{#2}}

\bibitem[{\citenamefont{Scully and Zubairy}(1997)}]{Scully97}
\bibinfo{author}{\bibfnamefont{M.~O.} \bibnamefont{Scully}} \bibnamefont{and}
  \bibinfo{author}{\bibfnamefont{M.}~\bibnamefont{Zubairy}},
  \emph{\bibinfo{title}{Quantum Optics}} (\bibinfo{publisher}{Cambridge
  University Press}, \bibinfo{address}{Cambridge, UK},
  \bibinfo{year}{1997}).

\bibitem[{\citenamefont{Kimble et~al.}(1977)\citenamefont{Kimble, Dagenais, and
  Mandel}}]{Kimble77}
\bibinfo{author}{\bibfnamefont{H.~J.} \bibnamefont{Kimble}},
  \bibinfo{author}{\bibfnamefont{M.}~\bibnamefont{Dagenais}}, \bibnamefont{and}
  \bibinfo{author}{\bibfnamefont{L.}~\bibnamefont{Mandel}},
  \bibinfo{journal}{Phys. Rev. Lett.} \textbf{\bibinfo{volume}{39}},
  \bibinfo{pages}{691} (\bibinfo{year}{1977}).

\bibitem[{\citenamefont{Carmichael et~al.}(1978)\citenamefont{Carmichael,
  Drummond, Meystre, and Walls}}]{Carmichael78}
\bibinfo{author}{\bibfnamefont{H.~J.} \bibnamefont{Carmichael}} et al.,
%  \bibinfo{author}{\bibfnamefont{P.}~\bibnamefont{Drummond}},
%  \bibinfo{author}{\bibfnamefont{P.}~\bibnamefont{Meystre}}, \bibnamefont{and}
%  \bibinfo{author}{\bibfnamefont{D.~F.} \bibnamefont{Walls}},
  \bibinfo{journal}{J. Phys. A} \textbf{\bibinfo{volume}{11}},
  \bibinfo{pages}{L121} (\bibinfo{year}{1978}).

\bibitem[{\citenamefont{Diedrich and Walther}(1987)}]{Diedrich87}
\bibinfo{author}{\bibfnamefont{F.}~\bibnamefont{Diedrich}} \bibnamefont{and}
  \bibinfo{author}{\bibfnamefont{H.}~\bibnamefont{Walther}},
  \bibinfo{journal}{Phys. Rev. Lett.} \textbf{\bibinfo{volume}{58}},
  \bibinfo{pages}{203} (\bibinfo{year}{1987}).

\bibitem[{\citenamefont{Gomer et~al.}(1998)\citenamefont{Gomer, Strauch,
  Ueberholz, Knappe, and Meschede}}]{Gomer98}
\bibinfo{author}{\bibfnamefont{V.}~\bibnamefont{Gomer}} et al.,
%  \bibinfo{author}{\bibfnamefont{F.}~\bibnamefont{Strauch}},
%  \bibinfo{author}{\bibfnamefont{B.}~\bibnamefont{Ueberholz}},
%  \bibinfo{author}{\bibfnamefont{S.}~\bibnamefont{Knappe}}, \bibnamefont{and}
%  \bibinfo{author}{\bibfnamefont{D.}~\bibnamefont{Meschede}},
  \bibinfo{journal}{Phys. Rev. A} \textbf{\bibinfo{volume}{58}},
  \bibinfo{pages}{R1657} (\bibinfo{year}{1998}).

\bibitem[{\citenamefont{McKeever et~al.}(2003)\citenamefont{McKeever, Boca,
  Boozer, Buck, and Kimble}}]{McKeever03:2}
\bibinfo{author}{\bibfnamefont{J.}~\bibnamefont{McKeever}} et al.,
%  \bibinfo{author}{\bibfnamefont{A.}~\bibnamefont{Boca}},
%  \bibinfo{author}{\bibfnamefont{A.~D.} \bibnamefont{Boozer}},
%  \bibinfo{author}{\bibfnamefont{J.~R.} \bibnamefont{Buck}}, \bibnamefont{and}
%  \bibinfo{author}{\bibfnamefont{H.~J.} \bibnamefont{Kimble}},
  \bibinfo{journal}{Nature} \textbf{\bibinfo{volume}{425}},
  \bibinfo{pages}{268} (\bibinfo{year}{2003}).

\bibitem[{\citenamefont{De Martini
  et~al.}(1996{\natexlab{a}})\citenamefont{De Martini, Di Giuseppe, and Marrocco}}]{Martini96}
\bibinfo{author}{\bibfnamefont{F.}~\bibnamefont{De Martini}},
\bibinfo{author}{\bibfnamefont{G.}~\bibnamefont{Di Giuseppe}},
\bibnamefont{and} \bibinfo{author}{\bibfnamefont{M.}
\bibnamefont{Marrocco}}, \bibinfo{journal}{Phys. Rev. Lett.}
  \textbf{\bibinfo{volume}{76}}, \bibinfo{pages}{900}
  (\bibinfo{year}{1996}).

\bibitem[{\citenamefont{Michler
  et~al.}(2000)\citenamefont{Michler, Imamoglu, Mason, Carson,
  Strouse, and Buratto}}]{Michler00}
\bibinfo{author}{\bibfnamefont{P.}~\bibnamefont{Michler}} et al.,
%  \bibinfo{author}{\bibfnamefont{A.}~\bibnamefont{Imamoglu}},
%  \bibinfo{author}{\bibfnamefont{M.~D.} \bibnamefont{Mason}},
%  \bibinfo{author}{\bibfnamefont{P.~J.} \bibnamefont{Carson}},
%  \bibinfo{author}{\bibfnamefont{G.~F.} \bibnamefont{Strouse}},
%  \bibnamefont{and} \bibinfo{author}{\bibfnamefont{S.~K.}
%  \bibnamefont{Buratto}}, 
\bibinfo{journal}{Nature}
  \textbf{\bibinfo{volume}{406}}, \bibinfo{pages}{968}
  (\bibinfo{year}{2000}{\natexlab{a}}).

\bibitem[{\citenamefont{Kurtsiefer et~al.}(2000)\citenamefont{Kurtsiefer,
  Mayer, Zarda, and Weinfurter}}]{Kurtsiefer00}
\bibinfo{author}{\bibfnamefont{C.}~\bibnamefont{Kurtsiefer}} et al.,
%  \bibinfo{author}{\bibfnamefont{S.}~\bibnamefont{Mayer}},
%  \bibinfo{author}{\bibfnamefont{P.}~\bibnamefont{Zarda}}, \bibnamefont{and}
%  \bibinfo{author}{\bibfnamefont{H.}~\bibnamefont{Weinfurter}},
  \bibinfo{journal}{Phys. Rev. Lett.} \textbf{\bibinfo{volume}{85}},
  \bibinfo{pages}{290} (\bibinfo{year}{2000}).

\bibitem[{\citenamefont{Beveratos et~al.}(2001)\citenamefont{Beveratos, Brouri,
  Gacoin, Poizat, and Grangier}}]{Beveratos01}
\bibinfo{author}{\bibfnamefont{A.}~\bibnamefont{Beveratos}} et al.,
%  \bibinfo{author}{\bibfnamefont{R.}~\bibnamefont{Brouri}},
%  \bibinfo{author}{\bibfnamefont{T.}~\bibnamefont{Gacoin}},
%  \bibinfo{author}{\bibfnamefont{J.-P.} \bibnamefont{Poizat}},
%  \bibnamefont{and} \bibinfo{author}{\bibfnamefont{P.}~\bibnamefont{Grangier}},
  \bibinfo{journal}{Phys. Rev. A} \textbf{\bibinfo{volume}{64}},
  \bibinfo{pages}{061802} (\bibinfo{year}{2001}).

\bibitem[{\citenamefont{Kim et~al.}(1999)\citenamefont{Kim, Benson, Kan, and
  Yamamoto}}]{Kim99}
\bibinfo{author}{\bibfnamefont{J.}~\bibnamefont{Kim}} et al.,
%  \bibinfo{author}{\bibfnamefont{O.}~\bibnamefont{Benson}},
%  \bibinfo{author}{\bibfnamefont{H.}~\bibnamefont{Kan}}, \bibnamefont{and}
%  \bibinfo{author}{\bibfnamefont{Y.}~\bibnamefont{Yamamoto}},
  \bibinfo{journal}{Nature} \textbf{\bibinfo{volume}{397}},
  \bibinfo{pages}{500} (\bibinfo{year}{1999}).

\bibitem[{\citenamefont{Lounis and Moerner}(2000)}]{Lounis00}
\bibinfo{author}{\bibfnamefont{B.}~\bibnamefont{Lounis}} \bibnamefont{and}
  \bibinfo{author}{\bibfnamefont{W.~E.} \bibnamefont{Moerner}},
  \bibinfo{journal}{Nature} \textbf{\bibinfo{volume}{407}},
  \bibinfo{pages}{491} (\bibinfo{year}{2000}).

\bibitem[{\citenamefont{Michler
  et~al.}(2000{\natexlab{b}})\citenamefont{Michler, Kiraz, Becher, Schoenfeld,
  Petroff, Zhang, Hu, and Imamoglu}}]{Michler00:2}
\bibinfo{author}{\bibfnamefont{P.}~\bibnamefont{Michler}} et al.,
%  \bibinfo{author}{\bibfnamefont{A.}~\bibnamefont{Kiraz}},
%  \bibinfo{author}{\bibfnamefont{C.}~\bibnamefont{Becher}},
%  \bibinfo{author}{\bibfnamefont{W.~V.} \bibnamefont{Schoenfeld}},
%  \bibinfo{author}{\bibfnamefont{P.~M.} \bibnamefont{Petroff}},
%  \bibinfo{author}{\bibfnamefont{L.}~\bibnamefont{Zhang}},
%  \bibinfo{author}{\bibfnamefont{E.}~\bibnamefont{Hu}}, \bibnamefont{and}
%  \bibinfo{author}{\bibfnamefont{A.}~\bibnamefont{Imamoglu}},
  \bibinfo{journal}{Science} \textbf{\bibinfo{volume}{290}},
  \bibinfo{pages}{2282} (\bibinfo{year}{2000}{\natexlab{b}}).

\bibitem[{\citenamefont{Santori et~al.}(2002)\citenamefont{Santori, Fattal,
  Vu\v{c}kovi\'{c}, Solomon, and Yamamoto}}]{Santori02}
\bibinfo{author}{\bibfnamefont{C.}~\bibnamefont{Santori}} et al.,
%  \bibinfo{author}{\bibfnamefont{D.}~\bibnamefont{Fattal}},
%  \bibinfo{author}{\bibfnamefont{J.}~\bibnamefont{Vu\v{c}kovi\'{c}}},
%  \bibinfo{author}{\bibfnamefont{G.~S.} \bibnamefont{Solomon}},
%  \bibnamefont{and} \bibinfo{author}{\bibfnamefont{Y.}~\bibnamefont{Yamamoto}},
  \bibinfo{journal}{Nature} \textbf{\bibinfo{volume}{419}},
  \bibinfo{pages}{594} (\bibinfo{year}{2002}).

\bibitem[{\citenamefont{Kuhn et~al.}(2002)\citenamefont{Kuhn, Hennrich, and
  Rempe}}]{Kuhn02}
\bibinfo{author}{\bibfnamefont{A.}~\bibnamefont{Kuhn}},
  \bibinfo{author}{\bibfnamefont{M.}~\bibnamefont{Hennrich}}, \bibnamefont{and}
  \bibinfo{author}{\bibfnamefont{G.}~\bibnamefont{Rempe}},
  \bibinfo{journal}{Phys. Rev. Lett.} \textbf{\bibinfo{volume}{89}},
  \bibinfo{pages}{067901} (\bibinfo{year}{2002}).

\bibitem[{\citenamefont{Beveratos et~al.}(2002)\citenamefont{Beveratos, Brouri,
  Garcoin, Villing, Poizat, and Grangier}}]{Beveratos02:2}
\bibinfo{author}{\bibfnamefont{A.}~\bibnamefont{Beveratos}} et al.,
%  \bibinfo{author}{\bibfnamefont{R.}~\bibnamefont{Brouri}},
%  \bibinfo{author}{\bibfnamefont{T.}~\bibnamefont{Garcoin}},
%  \bibinfo{author}{\bibfnamefont{A.}~\bibnamefont{Villing}},
%  \bibinfo{author}{\bibfnamefont{J.-P.} \bibnamefont{Poizat}},
%  \bibnamefont{and} \bibinfo{author}{\bibfnamefont{P.}~\bibnamefont{Grangier}},
  \bibinfo{journal}{Phys. Rev. Lett.} \textbf{\bibinfo{volume}{89}},
  \bibinfo{pages}{187901} (\bibinfo{year}{2002}).

\bibitem[{\citenamefont{McKeever et~al.}(2004)\citenamefont{McKeever, Boca,
  Boozer, Miller, Buck, Kuzmich, and Kimble}}]{McKeever04}
\bibinfo{author}{\bibfnamefont{J.}~\bibnamefont{McKeever}} et al.,
%  \bibinfo{author}{\bibfnamefont{A.}~\bibnamefont{Boca}},
%  \bibinfo{author}{\bibfnamefont{A.~D.} \bibnamefont{Boozer}},
%  \bibinfo{author}{\bibfnamefont{R.}~\bibnamefont{Miller}},
%  \bibinfo{author}{\bibfnamefont{J.~R.} \bibnamefont{Buck}},
%  \bibinfo{author}{\bibfnamefont{A.}~\bibnamefont{Kuzmich}}, \bibnamefont{and}
%  \bibinfo{author}{\bibfnamefont{H.~J.} \bibnamefont{Kimble}},
  \bibinfo{journal}{Science} \textbf{\bibinfo{volume}{303}},
  \bibinfo{pages}{1992} (\bibinfo{year}{2004}).

\bibitem[{\citenamefont{Hanbury-Brown and Twiss}(1956)}]{Hanbury56}
\bibinfo{author}{\bibfnamefont{R.}~\bibnamefont{Hanbury-Brown}}
  \bibnamefont{and} \bibinfo{author}{\bibfnamefont{R.~Q.} \bibnamefont{Twiss}},
  \bibinfo{journal}{Nature} \textbf{\bibinfo{volume}{178}},
  \bibinfo{pages}{1046} (\bibinfo{year}{1956}).

\bibitem[{\citenamefont{Jurczak et~al.}(1996)\citenamefont{Jurczak, Desruelle,
  Sengstock, Courtois, Westbrook, and Aspect}}]{Jurczak96}
\bibinfo{author}{\bibfnamefont{C.}~\bibnamefont{Jurczak}} et al.,
%  \bibinfo{author}{\bibfnamefont{B.}~\bibnamefont{Desruelle}},
%  \bibinfo{author}{\bibfnamefont{K.}~\bibnamefont{Sengstock}},
%  \bibinfo{author}{\bibfnamefont{J.-Y.} \bibnamefont{Courtois}},
%  \bibinfo{author}{\bibfnamefont{C.~I.} \bibnamefont{Westbrook}},
%  \bibnamefont{and} \bibinfo{author}{\bibfnamefont{A.}~\bibnamefont{Aspect}},
  \bibinfo{journal}{Phys. Rev. Lett.} \textbf{\bibinfo{volume}{77}},
  \bibinfo{pages}{1727} (\bibinfo{year}{1996}).

\bibitem[{\citenamefont{Bali et~al.}(1996)\citenamefont{Bali, Hoffmann,
  Sim{\'a}n, and Walker}}]{Bali96}
\bibinfo{author}{\bibfnamefont{S.}~\bibnamefont{Bali}} et al.,
%  \bibinfo{author}{\bibfnamefont{D.}~\bibnamefont{Hoffmann}},
%  \bibinfo{author}{\bibfnamefont{J.}~\bibnamefont{Sim{\'a}n}},
%  \bibnamefont{and} \bibinfo{author}{\bibfnamefont{T.}~\bibnamefont{Walker}},
  \bibinfo{journal}{Phys. Rev. A} \textbf{\bibinfo{volume}{53}},
  \bibinfo{pages}{3469} (\bibinfo{year}{1996}).

\bibitem[{\citenamefont{Narducci}(1996)}]{Narducci96}
\bibinfo{author}{\bibfnamefont{F.~A.} \bibnamefont{Narducci}}, Ph.D. thesis,
  \bibinfo{school}{University of Rochester},
  \bibinfo{address}{Rochester (NY)} (\bibinfo{year}{1996}).

\bibitem[{\citenamefont{Rempe et~al.}(1991)\citenamefont{Rempe, Thompson,
  Brecha, Lee, and Kimble}}]{Rempe91}
\bibinfo{author}{\bibfnamefont{G.}~\bibnamefont{Rempe}} et al.,
%  \bibinfo{author}{\bibfnamefont{R.~J.} \bibnamefont{Thompson}},
%  \bibinfo{author}{\bibfnamefont{R.~J.} \bibnamefont{Brecha}},
%  \bibinfo{author}{\bibfnamefont{W.~D.} \bibnamefont{Lee}}, \bibnamefont{and}
%  \bibinfo{author}{\bibfnamefont{H.~J.} \bibnamefont{Kimble}},
  \bibinfo{journal}{Phys. Rev. Lett.} \textbf{\bibinfo{volume}{67}},
  \bibinfo{pages}{1727} (\bibinfo{year}{1991}).

\bibitem[{\citenamefont{Mielke et~al.}(1998)\citenamefont{Mielke, Foster, and
  Orozco}}]{Mielke98}
\bibinfo{author}{\bibfnamefont{S.~L.} \bibnamefont{Mielke}},
  \bibinfo{author}{\bibfnamefont{G.~T.} \bibnamefont{Foster}},
  \bibnamefont{and} \bibinfo{author}{\bibfnamefont{L.~A.}
  \bibnamefont{Orozco}}, \bibinfo{journal}{Phys. Rev. Lett.}
  \textbf{\bibinfo{volume}{80}}, \bibinfo{pages}{3948} (\bibinfo{year}{1998}).

\bibitem[{\citenamefont{Smith et~al.}(2002)\citenamefont{Smith, Reiner, Orozco,
  Kuhr, and Wiseman}}]{Smith02}
\bibinfo{author}{\bibfnamefont{W.~P.} \bibnamefont{Smith}} et al.,
%  \bibinfo{author}{\bibfnamefont{J.~E.} \bibnamefont{Reiner}},
%  \bibinfo{author}{\bibfnamefont{L.~A.} \bibnamefont{Orozco}},
%  \bibinfo{author}{\bibfnamefont{S.}~\bibnamefont{Kuhr}}, \bibnamefont{and}
%  \bibinfo{author}{\bibfnamefont{H.~M.} \bibnamefont{Wiseman}},
  \bibinfo{journal}{Phys. Rev. Lett.} \textbf{\bibinfo{volume}{89}},
  \bibinfo{pages}{133601} (\bibinfo{year}{2002}).

\bibitem[{\citenamefont{Hennrich et~al.}(2000)\citenamefont{Hennrich, Legero,
  Kuhn, and Rempe}}]{Hennrich00}
\bibinfo{author}{\bibfnamefont{M.}~\bibnamefont{Hennrich}} et al.,
%  \bibinfo{author}{\bibfnamefont{T.}~\bibnamefont{Legero}},
%  \bibinfo{author}{\bibfnamefont{A.}~\bibnamefont{Kuhn}}, \bibnamefont{and}
%  \bibinfo{author}{\bibfnamefont{G.}~\bibnamefont{Rempe}},
  \bibinfo{journal}{Phys. Rev. Lett.} \textbf{\bibinfo{volume}{85}},
  \bibinfo{pages}{4872} (\bibinfo{year}{2000}).

\bibitem[{\citenamefont{Kuhn and Rempe}(2002)}]{Kuhn02:2}
\bibinfo{author}{\bibfnamefont{A.}~\bibnamefont{Kuhn}} \bibnamefont{and}
  \bibinfo{author}{\bibfnamefont{G.}~\bibnamefont{Rempe}}, in
  \emph{\bibinfo{booktitle}{Experimental Quantum Computation and Information}},
  edited by \bibinfo{editor}{\bibfnamefont{F.}~\bibnamefont{De~Martini}}
  \bibnamefont{and} \bibinfo{editor}{\bibfnamefont{C.}~\bibnamefont{Monroe}}
  (\bibinfo{publisher}{IOS-Press}, \bibinfo{year}{2002}), vol.
  \bibinfo{volume}{148}, pp. \bibinfo{pages}{37--66}.

\bibitem[{\citenamefont{Hennrich et~al.}(2003)\citenamefont{Hennrich, Kuhn, and
  Rempe}}]{Hennrich03}
\bibinfo{author}{\bibfnamefont{M.}~\bibnamefont{Hennrich}},
  \bibinfo{author}{\bibfnamefont{A.}~\bibnamefont{Kuhn}}, \bibnamefont{and}
  \bibinfo{author}{\bibfnamefont{G.}~\bibnamefont{Rempe}}, \bibinfo{journal}{J.
  Mod. Opt.} \textbf{\bibinfo{volume}{50}}, \bibinfo{pages}{936}
  (\bibinfo{year}{2003}).

\bibitem[{\citenamefont{Kuhn et~al.}(2003)\citenamefont{Kuhn, Hennrich, and
  Rempe}}]{Kuhn03}
\bibinfo{author}{\bibfnamefont{A.}~\bibnamefont{Kuhn}},
  \bibinfo{author}{\bibfnamefont{M.}~\bibnamefont{Hennrich}}, \bibnamefont{and}
  \bibinfo{author}{\bibfnamefont{G.}~\bibnamefont{Rempe}}, in
  \emph{\bibinfo{booktitle}{Quantum Information Processing}}, edited by
  \bibinfo{editor}{\bibfnamefont{T.}~\bibnamefont{Beth}} \bibnamefont{and}
  \bibinfo{editor}{\bibfnamefont{G.}~\bibnamefont{Leuchs}}
  (\bibinfo{publisher}{Wiley-VCH, Berlin}, \bibinfo{year}{2003}), pp.
  \bibinfo{pages}{182--195}.

\bibitem[{\citenamefont{Kimble et~al.}(1978)\citenamefont{Kimble, Dagenais, and
  Mandel}}]{Kimble78}
\bibinfo{author}{\bibfnamefont{H.~J.} \bibnamefont{Kimble}},
  \bibinfo{author}{\bibfnamefont{M.}~\bibnamefont{Dagenais}}, \bibnamefont{and}
  \bibinfo{author}{\bibfnamefont{L.}~\bibnamefont{Mandel}},
  \bibinfo{journal}{Phys. Rev. A} \textbf{\bibinfo{volume}{18}},
  \bibinfo{pages}{201} (\bibinfo{year}{1978}).

\bibitem[{\citenamefont{Carmichael and Walls}(1976)}]{Carmichael76}
\bibinfo{author}{\bibfnamefont{H.~J.} \bibnamefont{Carmichael}}
  \bibnamefont{and} \bibinfo{author}{\bibfnamefont{D.~F.} \bibnamefont{Walls}},
  \bibinfo{journal}{J. Phys. B: Atom. Molec. Phys.}
  \textbf{\bibinfo{volume}{9}}, \bibinfo{pages}{1199} (\bibinfo{year}{1976}).

\end{thebibliography}
\end{document}